\input{aipcheck}

\documentclass[
    ,final            
  ]
  {aipproc}

\layoutstyle{8x11single}

\begin{document}

\title{The Use of Weighting in Periodicity Searches in
All-Sky Monitor Data: Applications to the GLAST
LAT}


\author{Robin Corbet}{
  address={CRESST/USRA/GSFC, Code 662 NASA/GSFC, Greenbelt
Road, Greenbelt MD 20771}
}

\author{Richard Dubois}{
  address={Stanford Linear Accelerator Center,
Menlo Park, CA 94025}
}

\begin{abstract}
The light curves produced by all-sky monitors, such as the Rossi X-ray
Timing Explorer All-Sky Monitor and the Swift Burst Alert Telescope
(BAT), generally have non-uniform error bars. In searching for
periodic modulation in this type of data using power spectra it can be
important to use appropriate weighting of data points to achieve the
best sensitivity. It was recently
demonstrated that for Swift BAT data a simple weighting scheme can
actually sometimes reduce the sensitivity of the power spectrum
depending on source brightness. Instead, a modified weighting scheme,
based on the Cochran semi-weighted mean, gives improved results
independent of source brightness. We investigate the benefits of
weighting power spectra in period searches using simulated GLAST LAT
observations of $\gamma$-ray binaries.
\end{abstract}

\maketitle


\section{Introduction}

The light curves produced by many types of all-sky monitor do not have
uniform errors. For example, the pointing direction of the Burst Alert
Telescope (BAT) on Swift is determined by observations using the
narrow-field instruments also onboard Swift which are
primarily used to study gamma-ray bursts and their afterglows. BAT
observations of X-ray sources are thus generally obtained in an
unpredictable fashion. The All-Sky Monitor (ASM) on
board RXTE also suffers from similar problems. Even though the RXTE
ASM is controlled to make observations of the sky as complete as
possible, there is still considerable non-uniformity. The quality of
observations of any particular point on the sky will depend on
the observation duration and the location
of that point in a detector field of view if there is non-uniform
response across the FOV.

\section{Weighted and Semi-Weighted Power Spectra}

Scargle (1989) proposed that the effect of unequally weighted data
points in a power spectrum can be found by considering
two points that coincide
and treating this as a single point of double weight. A ``natural'' 
approach to combining data points of different error bar size is to
use the weighted mean. This approach can be considered as calculating
the power spectrum of $y_i/\sigma_i^2$.
Although this procedure is very effective for faint sources,
if the scatter in data values is large compared to the
error bar sizes
weighting by error bars can be inappropriate.
This is because the concept of combining ``coincident'' points
must be treated carefully. In a power spectrum
it is not just points at the same time, but points at the
same phase, which are combined.
If there is a strong signal then, at all frequencies except
the modulation frequency, we will be combining points with
discrepant values and hence the weighted mean will
not be valid.

A refinement is to
treat source variability as an additional ``error''\cite{Corbet07a},
i.e. calculate the power spectrum of
$ \frac{y_i}{((f \sigma_i)^2 + V_S)}$ where
$f$  is a correction to nominal error bar size and
$V_S$ is the estimated variance due to source variability.
This procedure is related to the semi-weighted
mean,
\cite{Cochran37,Cochran54}
and hence may be termed ``semi-weighting''\cite{Corbet07b}.
Semi-weighting works well for sources
across a wide range of brightness, providing
that the correction factor $f$ is applied, and
gives improvements for
Swift BAT\cite{Corbet07a}, RXTE ASM and CGRO BATSE
light curves (Fig. 1).

\section{Analysis of Simulated GLAST LAT Light Curves}
The GLAST LAT\cite{Michelson03} is also a large field of
view detector, and it will provide a sensitive survey of
the $\gamma$-ray sky.
It is expected that GLAST will predominantly operate in a sky
survey mode which will give much more uniform sky coverage
than with the RXTE ASM or Swift BAT. However, precession
of GLAST's orbit will affect the exposure of each point on
the sky.
In order for weighting to improve period searches two conditions must
be met: the signal amplitude should not be very much larger than
typical error bar size, and the sizes of the errors on data points in
light curves must have considerable variation.  If these conditions
are not met then, although weighting may still be used, it will not
have a significant benefit.
To investigate whether weighting will be beneficial for
GLAST we used simulated light curves of $\gamma$-ray binaries that
have been presented elsewhere\cite{Dubois06} with
1 day or 90 minute binning.
For light curves accumulated in 1 day bins, sky coverage was
sufficiently uniform that weighting gave no improvement.

The 90 minute time bins were not synchronized to GLAST's orbital
period which produced large
changes in exposure from bin to bin.
For these bins the number of photons was low
and we calculated asymmetric error bars based on
Pearson's $\chi^2$\cite{Heinrich03}.
We investigated several symmetrization schemes to convert
error bars to a single weighting factor\cite{Audi97}.
It was found that weighting gave very strong benefits
for period detection. Of the symmetrization methods investigated,
the use of just the larger (upper) error bar on data points
gave the strongest gain in signal detection.
However, no conversion method gave results that could
be used in semi-weighting as the predicted source variability
variance values were not well defined.


\begin{figure}
  \includegraphics[height=.425\textheight,angle=-90]{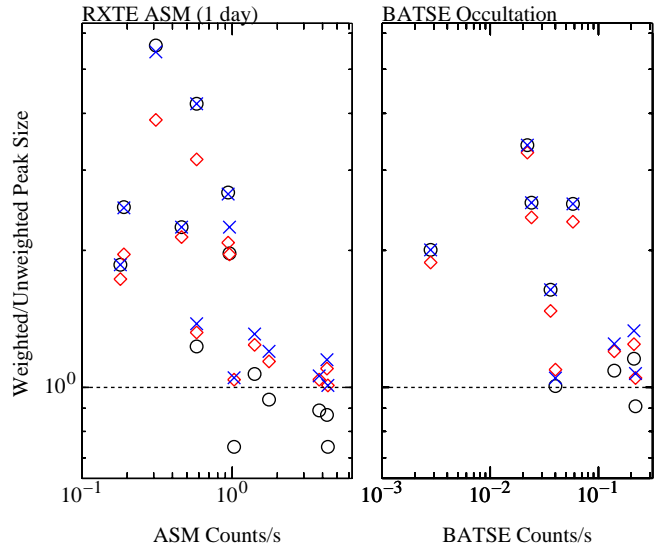}
  \caption{Comparison of power spectrum weighting techniques
for power spectra of RXTE ASM (left) and
BATSE Occultation (right) light
curves. The y-axes show the ratio of peak heights in weighted 
power spectra
compared to the heights in unweighted power
spectra. Black circles show simple weighting, blue crosses
semi-weighting, and red diamonds semi-weighting without error
bar correction applied. Points below the horizontal dashed
line indicate that weighting was worse than not weighting.}
\end{figure}

%


%



\bibliographystyle{aipprocl} 




\end{document}